\documentclass[aps,prl,preprint,superscriptaddress,letterpaper]{revtex4}

\usepackage{tikz}

\usepackage{graphicx}
\usepackage{dcolumn}
\usepackage{bm}

\newcolumntype{.}{D{.}{.}{8}}

\usepackage[utf8]{inputenc}
\usepackage{amsmath, amssymb}
\usepackage{tabularx,booktabs,array,dcolumn}
\usepackage{setspace}
\usepackage{vmargin}
\usepackage{xcolor}
\usepackage{graphicx,psfrag,subfigure}

\usepackage{float}
\usepackage[all]{xy}

\usepackage{mathtools}
\usepackage[english]{babel}

\newcommand{\bos}[1]{\boldsymbol{#1}}
\newcommand{\mr}[1]{\mathrm{#1}}
\newcommand{\pd}[2]{\frac{\partial #1}{\partial #2}}

\def\Eh{$\text{E}_\text{h}$}
\def\tr{^\mathrm{T}}
\def\iim{\mr{i}}
\def\eem{\mr{e}}

\def\DC{\text{DC}}
\def\DCB{\text{DCB}}

\def\eplus{E_+}
\def\bPsi{\bos{\Psi}}
\def\np{n} 
\def\nnuc{N} 
\def\nb{N_\text{b}} 
\def\bare{\text{bare}}

\def\unittwo{1^{[2]}} 
\def\unitfour{1^{[4]}} 
\def\unitthree{1^{[3]}} 
\def\unitsixteen{1^{[16]}} 
\def\unitfull{1^{[4^{\np}]}} 

\def\four{^{[4]}} 
\def\sixteen{^{[16]}} 
\def\Xfour{X^{[4]}} 

\def\ll{\mathrm{ll}}
\def\ls{\mathrm{ls}}
\def\sl{\mathrm{sl}}
\def\ss{\mathrm{ss}}

\def\uu{\mathrm{\uparrow\uparrow}}
\def\ud{\mathrm{\uparrow\downarrow}}
\def\du{\mathrm{\downarrow\uparrow}}
\def\dd{\mathrm{\downarrow\downarrow}}

\def\varphip{\varphi_0^{(+)}}

\def\bp{\bos{p}}
\def\br{\bos{r}}
\def\balpha{\bos{\alpha}}
\def\bsigma{\bos{\sigma}}

\def\bB{{K}_\text{B}}
\def\vtheta{\vartheta}

\def\som{Supplementary Material}

\begin{document}

\title{%
All-order relativistic computations for atoms and molecules
using an explicitly correlated Gaussian basis
}

\author{P\'eter Jeszenszki} 
\author{D\'avid Ferenc} 
\author{Edit M\'atyus} 
\email{edit.matyus@ttk.elte.hu}
\affiliation{Institute of Chemistry, ELTE, Eötvös Loránd University, 
Pázmány Péter sétány 1/A, Budapest, H-1117, Hungary}

\date{\today}

\begin{abstract}
\noindent %
A variational solution procedure is reported for 
the many-particle no-pair Dirac--Coulomb--Breit Hamiltonian
aiming at a parts-per-billion (ppb) convergence of 
the atomic and molecular energies, described within the fixed nuclei approximation. 
The procedure is tested 
for nuclear charge numbers from $Z=1$ (hydrogen) to $28$ (iron). 
Already for the lowest $Z$ values, 
a significant difference is observed from 
leading-order Foldy--Woythusen perturbation theory, but 
the observed deviations are smaller than the estimated self-energy and 
vacuum polarization corrections.
\end{abstract}

\maketitle
%
\noindent %
Precision spectroscopy experiments carried out for small atomic \cite{hatom13,protonsize17,GuBaHoCa20} and molecular \cite{AlHaKoSc18,HoBeMe19} systems
have been proposed as low-energy tests of the fundamental theory of matter \cite{rmp18}. 
Atoms and molecules are bound many-body quantum systems held together by electromagnetic interactions usually complemented with some model for the internal nuclear structure.
Relativistic quantum electrodynamics is a simple $U(1)$ gauge theory with a Lagrangian density that, of course, obeys Lorentz invariance of special relativity that is standard textbook material \cite{KakuBook93}. 

At the same time, the bound states of atoms and molecules are conveniently obtained as eigenstates of some wave equation, most commonly as stationary states of 
the Galilean invariant Schr\"odinger equation.
Sophisticated techniques have been developed 
for a numerically exact (14 digit) solution of the three- \cite{Ko18h2p} and four-body 
Schrödinger equation \cite{PaKo18four}. `Effects' 
due to special relativity and the quantized fermion and photon fields
are accounted for as perturbation following and considerably extending 
the pioneer perturbation theory work that was first summarized in a book by Bethe and Salpeter in 1957 \cite{BeSabook57}. 
Further progress in this direction of research
is nowadays called the non-relativistic quantum electrodynamics (nrQED) approach \cite{LepagePhD1978,Eides2001,Pa06,HaZhKoKa20}
and it is successfully used for light atoms and molecules in comparison with precision spectroscopy experiments. 
There are sophisticated methods developed for the numerically stable evaluation 
of the increasingly complex correction formulae \cite{Pa06,KoHiKa13,PuKoCzPa16}
of the nrQED series expanded in terms of the $\alpha$ fine structure constant.

A practical, fully Lorentz covariant wave equation for many-spin-1/2 fermion systems is unknown, except for the two-particle case, for which the Bethe--Salpeter (BS) equation \cite{BeSa51} (see also Ref.~\cite{Na50}) 
offers a quantum-electrodynamics wave 
equation by properly accounting for also the relative time of the particles. 
Beyond two particles, formulation of a practical (and fully Lorentz covariant) QED wave equation remains to be a challenging problem \cite{JaPa21}. 
Following these observations,  the Galilean Schrödinger wave equation may appear to be a solid starting point for describing the molecular regime combined with the nrQED perturbative scheme that can be related
to a perturbative calculation of level shifts from the poles of the QED Green's function \cite{GMLo51,Su57,Mo89}.

At the same time, the Schrödinger wave equation is known to be an inaccurate
starting approximation for atoms and molecules, especially for nuclei beyond the lowest $Z$ nuclear charge numbers \cite{KeFaBook07,ReWoBook15}. 
Therefore, a `hybrid model' has been adopted in the quantum chemistry practice with assuming 
equal times for the particles but using Dirac's kinetic energy operator for every electron and describing the electron-electron interaction within some (most commonly the Coulomb) approximation:
\begin{align}
  {\tilde H}^{\bare}
  =
  \sum_{i=1}^\np
  h_{i}  
  + 
  \sum_{i=1}^\np u_{i}
  + 
  \sum_{i=1}^\np \sum_{j>i}^\np
  v_{ij}
  \label{eq:Hbare}
\end{align}
with $h_{i}=\unitfour(1)\otimes \ldots\otimes h\four_i(i) \otimes \ldots \otimes \unitfour(\np)$
and $h\four_i= c \balpha\four \cdot \bp + \beta\four m_i c^2$, 
where $\alpha_i$ and $\beta$ are the standard Dirac matrices.
The corresponding wave equation is neither fully Galilean, nor Lorentz invariant, 
but it should serve as a better
starting point than the Schrödinger equation. In particular, it
would allow to account
for the relativistic `effects' on an equal footing with 
electron correlation that is important for a good description of the molecular regime.
This \emph{ad hoc} construction has mathematical problems due to the non-positive definiteness of the operator. 

Sucher proposed \cite{Su80,Su83,HaSu84,Su84} a no-pair many-particle Hamiltonian based on relativistic QED 
that is reminiscent to the \emph{na\"ively} constructed Hamiltonian 
in Eq.~(\ref{eq:Hbare}) with the important difference that it is projected with $\Lambda_+$
to the positive energy states ($\eplus$) of a non-interacting reference problem,
$H_0=\sum_{i=1}^\np (h_{i}+u_{i})$:
\begin{align}
  {H}
  =
  \sum_{i=1}^N
  \Lambda_+ (h_{i}+ u_{i}) \Lambda_+ 
  + 
  \sum_{i=1}^N \sum_{j>i}^N
  \Lambda_+ v_{ij} \Lambda_+  \;.
  \label{eq:dirac}
\end{align}
Sucher explains \cite{Su80,Su83,HaSu84,Su84} that $H_0$ can be either the kinetic energy of 
the free spin-1/2 fermions or some other bound model without fermion-fermion interactions 
following Furry's work \cite{Fu51}. 
This no-pair operator, in some cases called the Brown--Ravenhall operator \cite{HoSi99}, 
has well-defined mathematical properties, and most importantly, it is bounded from below.
The `no-pair' expression refers to the fact that, due to the $\Lambda_+$ projection, 
this Hamiltonian does not account for pair creation of the spin-1/2 particles (\emph{e.g.,} electron-positron pairs) of the $H_0$ non-interacting model, but it operates with a fixed fermion number. 
This is a natural starting point for describing chemical systems. 
Pair effects can be accounted for in a next stage of the theoretical treatment.

The present work is about the development and application of a practical variational
procedure for solving the
\begin{align}
  H
  \bPsi
  =
  E\bPsi
\end{align}
wave equation for atoms and also molecules
on the order of a parts-per-billion (ppb) precision.
This development is an important step towards 
providing benchmark theoretical values for precision spectroscopy experiments
and also an independent test for the nrQED computations.
In  further work, we plan to account for the effect of pair creation 
and for interaction with the photon modes that is necessary
for a direct comparison.

There is already, of course, important work in the literature about precise 
variational relativistic approaches for atoms. 
Grant and co-workers developed the GRASP computer program to treat atoms especially with high $Z$-values  
\cite{PaGr90,PaFFGr96} starting out from the Dirac--Hartree--Fock (Dirac HF) framework. 
Shabaev and co-workers have reported several developments for atoms 
also based on the Dirac HF model as a starting reference. 
They developed the QED model operator approach \cite{ArShYePlSo05,ShTuYe13,ShTuKaKoMaMi20}
for computing self-energy corrections, 
and their most recent applications include results for resonance states of medium $Z$ heliumlike ions \cite{ZaMaTuSh19} including the exact one-photon exchange, pair creation, and self-energy corrections.
Benchmark results were reported by Bylicki, Pestka, and Karwowski \cite{ByPeKa08} for two-electron atoms using the Dirac--Coulomb operator, a $\Lambda_+$ projector similar to ours, and an explicitly correlated Hylleraas basis set.

Regarding the molecular regime, it is necessary to mention the BERTHA 
\cite{QuSkGr98}
and the DIRAC \cite{DIRAC19}
program packages that include implementation of hierarchical quantum chemistry methods starting with the HF approximation and typically aiming for chemical accuracy in the computational results. 

For the present work, we restrict the discussion to two spin-1/2 fermions
and the fixed nuclei (`Born--Oppenheimer') approximation. 
The restriction on the number of particles can be lifted without conceptual difficulties and 
we can foresee applications (with the ppb convergence criterion) to 3-4 particles.
It should also be possible to include spin-1/2 nuclei in the treatment 
on the same footing as the electrons \cite{MaRe12,Ma13,FeMa19a,Ma19review}
(first assuming point-like, structureless nuclei as if they were elementary spin-1/2 particles). In this case, it appears to be a natural choice to use 
(the finite basis representation) of the free-particle projector, or 
to explore some other possible non-interacting reference system 
specifically designed for the pre-Born--Oppenheimer problem.

For the present description of atoms and molecules with clamped nuclei, 
it is a natural choice for the definition of the $\Lambda_+$ projector to use the non-interacting two-electron model that is bound by the external potential of the fixed nuclei (without electron-electron interaction).
In our implementation, we can work with other non-interacting models
to define the projector, including the finite-basis free-electron model
or other external field one-electron systems. 
It remains a question to be explored in future work, which choice will be the most convenient
one for further numerical applications, and in particular, for the incorporation
of (electron-positron and photon) field interactions.

In this work, we build the $\Lambda_+=\sum_n |\varphi^{(+)}_{0,n} \rangle \langle \varphi^{(+)}_{0,n}|$ 
projector from the $\varphi^{(+)}_{0,n}$ 
eigenstates of the atomic or molecular Hamiltonian without electron-electron interactions that have positive energy ($E_+$) and do not belong to the Brown--Ravenhall (BR) continuum (that is uncoupled from the physical $E_+$ states in the absence of electron-electron interactions) \cite{BrRa51,PeByKa07,Ka17}. 
The physically relevant $E_+$ states are separated in practice from the BR states using the complex coordinate rotation (CCR) technique following Ref.~\cite{ByPeKa08}. 
The non-interacting computation is carried out
with the same basis set as the interacting computation, because the aim is 
to select (construct) the $E_+$ part of the actual basis for the full (interacting) problem.

The no-pair Hamiltonian including the fermion-fermion interactions reads as 
\begin{align}
  &H(1,2,\ldots,n) =
  \nonumber \\
  &=
  \Lambda_+ 
  \left\lbrace%
    \sum_{i=1}^\np
      \unitfour(1)\boxtimes \ldots\boxtimes  \hat{h}\four_i(i) 
      \boxtimes \ldots \boxtimes \unitfour(\np) 
    +
    \sum_{i=1}^\np 
        u_{i} \unitfull 
  \right.
    \nonumber \\
  &\quad +   
  \left.
  \sum_{i=1}^\np \sum_{j>i}^\np
    \left[%
      v_{ij} \unitfull + 
      x_{ij} \unitfour(1)\boxtimes \ldots\boxtimes  
      \balpha\four(i) \boxtimes \ldots  \boxtimes 
      \balpha\four(j) \boxtimes \ldots \unitfour(\np)
    \right]
  \right\rbrace
  \Lambda_+
  \label{eq:HTS} \; ,
\end{align}
where we used the $\boxtimes$ Tracy--Singh, `block-wise' direct product \cite{TrSi72,SiMaRe15},
that allows us to work with the Pauli's $\sigma$ matrices  and the `large' and `small' component blocks of the Dirac spinors.

In particular, the Hamiltonian operator for two spin-1/2 fermions (for convenience, shifted by $2m_ic^2$ for both $i=1$ and 2) takes
the following matrix form 
\begin{align}
  &H(1,2) =
  \nonumber \\
  &
  {\footnotesize
  \Lambda_+
  \left(%
    \begin{array}{@{} c@{}c@{}c@{}c @{}}
       V\unitfour+U \unitfour & 
       c \bsigma\four_2 \cdot \bp_2 & 
       c \bsigma\four_1 \cdot \bp_1 & 
       \Xfour \\
       c\bsigma\four_2 \cdot \bp_2 & 
       V\unitfour+(U - 2m_2c^2)\unitfour & 
       \Xfour & 
       c \bsigma\four_1 \cdot \bp_1 \\
       c\bsigma\four_1 \cdot\bp_1 & 
       \Xfour &
       V\unitfour+(U-2m_1c^2)\unitfour & 
       c \bsigma\four_2 \cdot \bp_2 \\
       \Xfour & 
       c \bsigma\four_1 \cdot \bp_1 &
       c \bsigma\four_2 \cdot \bp_2 & 
       V\unitfour+(U-2m_{12}c^2)\unitfour \\
    \end{array}
  \right)
  \Lambda_+
  }
  \label{eq:fullHam}
\end{align}
with $m_{12}=m_1+m_2$, $\bp_i = -\iim(\pd{}{r_{ix}},\pd{}{r_{iy}},\pd{}{r_{iz}})$ ($i=1,2$),
$\bsigma\four_1=(\sigma_x\otimes\unittwo,\sigma_y\otimes\unittwo,\sigma_z\otimes\unittwo)$
and
$\bsigma\four_2=(\unittwo\otimes \sigma_x,\unittwo\otimes\sigma_y,\unittwo\otimes\sigma_z)$, where $\sigma_x,\sigma_y,$ and $\sigma_z$ are the $2\times 2$ Pauli matrices.
Interactions with the fixed external electric charges (clamped nuclei) 
are collected in  $U=\sum_{i=1}^n \sum_{a=1}^{\nnuc}q_iQ_a/|r_i-R_a|$. 

The electron-electron interaction appears in the 
$4\times 4$ dimensional $V\unitfour$ and $\Xfour$ blocks. 
Regarding the matrix representation of the Hamiltonian
in the non-interacting two-electron basis,
the diagonal $16\times 16$ dimensional blocks 
(identical `in' and `out' energies) contain the one-photon exchange terms in leading-order, 
whereas the off-diagonal $16\times 16$ blocks (different `in' and `out' energies) 
assume photon emission or absorption, hence correspond to a process involving at least two photons.
In the present work, we describe the electron-electron interaction expressed
in the Coulomb gauge and invoke the zero-frequency approximation ($\omega\approx 0$) 
that gives rise to the Coulomb--Breit (CB) interaction operator. 
Within this approximation $H(1,2)$ is the no-pair Dirac--Coulomb--Breit (DCB) Hamiltonian 
that accounts for retardation to leading order, and
\begin{align}
  V
  &=
  \frac{q_1 q_2}{r_{12}} \quad\text{and}\\
  X\four
  &=  
  -
  q_1q_2
  \left[%
    \frac{\bos{\sigma}_1\bos{\sigma}_2}{r_{12}} 
    +\frac{1}{2} 
    (\bos{\sigma}_1\cdot\bos{\nabla}_1)
    (\bos{\sigma}_2\cdot\bos{\nabla}_2) r_{12}
  \right]   \; .
\end{align}
If $X\four$ is neglected ($X\four= 0\four$), we obtain the Dirac--Coulomb (DC) approximation that corresponds to instantaneous 
interactions.
We note that since both the Coulomb and the Coulomb--Breit approximations are independent 
of the frequency of the exchanged photons, they can be defined without explicit reference 
to the underlying `non-interacting' $\varphip$ representation.
Thus, the $\Lambda_+$ projection amounts to simple matrix multiplication with 
the `bare' (CCR scaled) Dirac Hamiltonian. 

To build the matrix representation of the two-electron Hamiltonian, we consider the wave function as a linear combination of 16-dimensional spinor basis functions, $\Psi^{\bos{\lambda}}_{i,\bos{\varsigma}} = \Psi^{\lambda_1,\lambda_2}_{i,\varsigma_1,\varsigma_2}(\br_1,\br_2)$:
\begin{align}
  \Psi(\br_1,\br_2) 
  =
  \sum_{i=1}^{\nb}
  \sum_{\bos{\lambda}=\lbrace{\ll,\ls,\sl,\ss\rbrace}} 
  \sum_{\bos{\varsigma}=\lbrace{\uu,\ud,\du,\dd\rbrace}}
    c_{i,\bos{\lambda},\bos{\varsigma}} 
    \Psi^{\bos{\lambda}}_{i,\bos{\varsigma}}(\br_1,\br_2) \;.
    \end{align}
For two identical spin-1/2 fermions, it is necessary to antisymmetrize the spinor basis (now, collecting the $\lambda_1\lambda_2$ blocks and the $\varsigma_1\varsigma_2$ spin components into one vector) that reads as
\begin{align}
  \Psi_i(\br_1,\br_2) 
  =
  \left\lbrace%
    \unitsixteen - \Pi\sixteen
  \right\rbrace
  \Phi_i(\br_1,\br_2) \; ,
  \label{eq:antisymecg}
\end{align}
where $\Pi\sixteen=P\four_{\text{ls}} \otimes P\four_{\uparrow \downarrow} P_{12} $
with the
$P\four_{\ls} = P \four_{\ud} =
((1,0,0,0),(0,0,1,0),(0,1,0,0),(0,0,0,1))$ matrices and $P_{12}$ is the coordinate exchange operator.
Furthermore, it is necessary to ensure spatial symmetry relations between the large (l) and the small (s) components in a finite basis representation of the Dirac operator. 
To represent the $(\bsigma\cdot\bp)(\bsigma\cdot\bp)=p^2$ identity in the finite spinor basis, we use the simplest two-particle kinetic balance (KB) condition \cite{Ku84,SiMaRe15}
of the large and small components:
\begin{align}
  &\left(%
  \begin{array}{@{}c@{}}
    \phi^\ll \\
    \phi^\ls \\
    \phi^\sl \\
    \phi^\ss \\
  \end{array}
  \right)
  = 
  \bB\sixteen \left(\begin{array}{c} 
       \Theta  \\
       \Theta \\
       \Theta \\
       \Theta
  \end{array} \right) 
  \quad \text{with}\quad
  &\bB\sixteen = \left( \begin{array}{cccc}
       \unitfour  & 0\four & 0\four & 0\four \\
       0\four &  \frac{\bsigma\four_2\cdot\bp_2}{2m_2c} & 0\four & 0\four  \\
       0\four & 0\four & \frac{\bsigma\four_1\cdot\bp_1}{2m_1c} & 0\four \\
       0\four & 0\four & 0\four & \frac{\bsigma\four_1\cdot\bp_1 \bsigma\four_2\cdot\bp_2}{4m_1m_2c^2}
  \end{array} \right)  \label{kbmetric}
\end{align}
that allows us to generate the $\lambda_1\lambda_2=$(ll,ls,sl,ss) blocks from the same 
four-dimensional $\Theta$ vector in which each element contains the same spatial function $\vtheta$,
$\Theta\tr=(\vartheta,\vartheta,\vartheta,\vartheta)$. 
For the $\vtheta$ spatial basis functions, we use floating explicitly correlated Gaussians (ECGs)
\begin{align}
  \vtheta_i
  =
  \eem^{-(r-s_i)\tr (\bos{A}_i\otimes \unitthree)  (r-s_i)}
  \label{eq:ecg}
\end{align}
that allow for an efficient description of the particle (electron) correlation \cite{rmp13},
and $\bos{A}_i\in\mathbb{R}^{\np\times\np}$ (symmetric, positive definite) and $\bos{s}_i\in\mathbb{R}^{3\np}$
are parameters optimized by minimization of the energy.

After considering the antisymmetrization
and kinetic balance equations, Eqs.~(\ref{eq:antisymecg}) and (\ref{kbmetric}), 
a $16\times 16$ dimensional block of the Hamiltonian and overlap matrices can be written as
\begin{align}
  \langle
    \Phi_{i}
    | O|
    \Phi_{j}
  \rangle 
  =
  \langle \Theta_i | \bB^\dagger O \bB | \Theta_j \rangle -  
  \langle \Theta_i | \bB^\dagger \Pi O \bB | \Theta_j \rangle \;, \quad O=H\ \text{or}\ I\; ,
\end{align}
for which we have calculated the analytic matrix elements with 
ECGs, Eq.~(\ref{eq:ecg}), and implemented the integral expressions in QUANTEN \cite{quanten}. 
We obtain the ground state as the lowest-energy (real) eigenvalue of the
generalized eigenvalue problem:
$ \bos{H} \bos{c} = E \bos{S} \bos{c}$
as an upper bound to the exact no-pair energy.

A good starting basis parameterization, Eq.~(\ref{eq:ecg}), 
for the systems studied in this work 
was obtained by minimization of the non-relativistic energy
to a ppb precision for the largest basis set sizes.
The numerical uncertainty of the values reported in this paper 
is determined by the double precision (8-byte reals)
arithmetics used in the optimization procedure.

In what follows, we report ground (and one example for excited) state energies 
obtained in the variational procedure implemented in the QUANTEN computer program and 
using the no-pair DC and
DCB Hamiltonians. In all computations, we used the CODATA18 value for the inverse fine-structure 
constant $\alpha^{-1}=137.035\ 999\ 084$ \cite{codata18}.

\begin{figure}
  \centering
  \scalebox{0.70}{\includegraphics{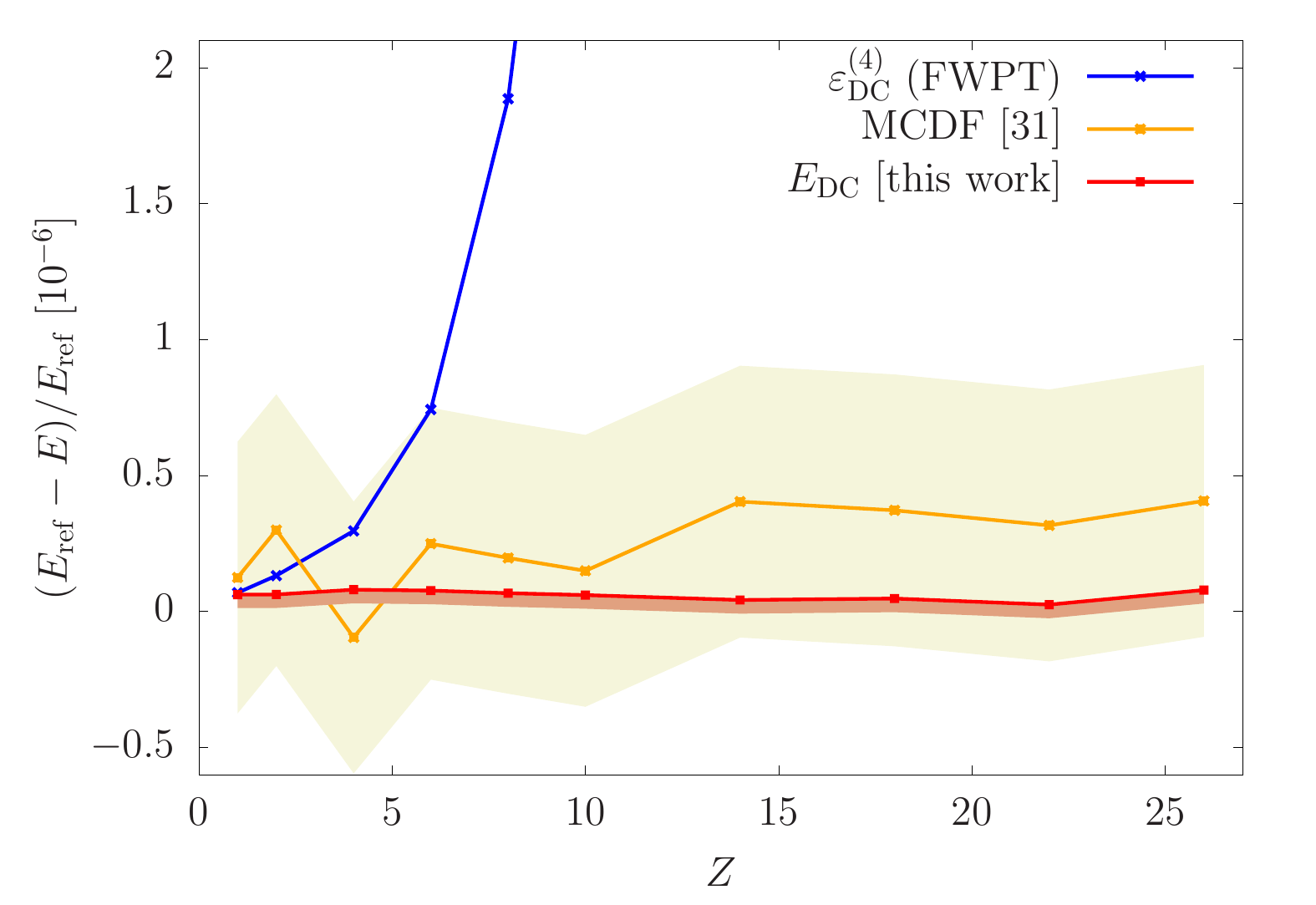}}
      \caption{%
        Dirac--Coulomb energy of two-electron helium-like ions. 
        Leading-order perturbative relativistic energy, $\epsilon^{(4)}_\text{DC}$ (for $Z=1$--$4$ \cite{Dr06}, for $4<Z\leq 26$ this work), extrapolated MCDF result \cite{PaGr90}; and the no-pair, variational $E_\text{DC}$ energy obtained in the present work. 
        $E_\text{ref}$ is the no-pair variational DC energy in a large Hylleraas basis \cite{ByPeKa08}.
        \label{fig:DCprojZ}
      }
\end{figure}

Figure~\ref{fig:DCprojZ} shows the excellent agreement of the atomic no-pair DC energies
with $Z=1$--26 nuclear charge numbers 
obtained in our implementation with benchmark literature data:
basis set extrapolated multi-configuration Dirac--Fock (MCDF) energies computed by 
Parpia and Grant \cite{PaGr90} and 
the DC energies of Bylicki, Pestka, and Karwowski obtained with a large Hylleraas basis set \cite{ByPeKa08}.
Using 300--400 ECGs, we observe an at least 8 digit agreement for $Z=1$--26 with Bylicki et al. 
The 30-year old extrapolated DC results (corresponding to an implicit HF projection) of 
Parpia and Grant perform remarkably well over the entire range,
but they have larger error bounds than Ref.~\cite{ByPeKa08} or our work. 
Foldy--Woythusen perturbation theory (FWPT) shows a deviation from these results that grows rapidly with $Z$. 
Parpia and Grant reported also the perturbative correction due to the exact one-photon exchange
to their MCDF wave function (with large error bounds). 
Comparison of their work with our no-pair DCB result
as well as with the leading-order FWPT DCB energy, $\epsilon_\DCB^{(4)}$ (expectation value of the Breit--Pauli Hamiltonian) is provided in the \som.

Since no variational reference data (of similar precision) is available for molecular systems, 
we will compare our results with FWPT energies that are known to high precision. 
Table~\ref{tab:all} summarizes numerical results
for the ground-state electronic energy 
of H$_2$, HeH$^+$, and H$_3^+$ with nuclei fixed close to the equilibrium structure
(convergence details are provided in the \som).
Due to the surprisingly large deviation of the leading-order FWPT energies 
($m\alpha^4$, $\epsilon^{(4)}$) and our variational values, we have considered also 
higher-order FWPT energies within the nrQED framework ($m\alpha^6$, $\epsilon^{(6)}$).
Regarding the `poly-electronic' systems, 
the involved computation of $\epsilon^{(6)}$ has been carried out so far
only for the H$_2$ molecule \cite{PuKoCzPa16}. 
For a better comparison, we include results in the table also for the ground
and the first excited singlet states of the He atom, the other `poly-electronic' system for which $\epsilon^{(6)}$ energies are available \cite{Pa06}.

Regarding the electronic ground state of the H$_2$ and H$_3^+$ hydrogenic compounds,  
our no-pair variational DC and DCB energy is
lower than the leading-order FWPT energy by 21-24~n\Eh\ and 58-90~n\Eh, respectively. 
For the ground state of the HeH$^+$ molecular ion and the He atom, 
our variational DC and DCB energies are lower by 142-146 and 638-712~n\Eh\ 
than the leading-order FWPT energy. 
It is interesting to note that these deviations are one-two orders of magnitude larger than 
the deviation of the exact and perturbative relativistic energy 
of the ground state, one-electron atomic hydrogen (0.18~n\Eh) and hydrogen-like helium (11~n\Eh) \cite{Eides2001}. 

Higher-order ($m\alpha^6$, $\epsilon^{(6)}$) FWPT results 
can be interpreted within the nrQED framework \cite{Pa06,PuKoCzPa16}. 
In the $m\alpha^6$ expressions of nrQED, it is possible to identify the higher (second) order 
FWPT correction corresponding to the DCB Hamiltonian. It turns out that
this correction contains divergent terms 
(due to the internal mathematical structure of the nrQED expansion). 
These divergent terms are cancelled with other divergent terms 
in the one- and two-photon exchange (also approximated at the $m\alpha^6$ level and expanded
with respect to the non-relativistic reference state) in Refs.~\cite{Pa06,PuKoCzPa16}. 
So, in the present comparison, we include the higher-order FWPT corrections
for the DCB operator after the divergences are cancelled within the nrQED expansion. 
The resulting $\epsilon^{(6)}$ energies are a little bit closer to 
the variational DCB result than the $\epsilon^{(4)}$ values, but 
the observed deviation remains large, 
$-56$~n\Eh\ and $-659$~n\Eh\ 
for the ground state of H$_2$ and He, respectively.
It is interesting to note that for the excited 1s2s ($2\ ^1$S) state of the helium atom, 
the agreement of the variational and FWPT energies is much better, 
and it significantly improves upon inclusion of higher-order PT corrections,
the $-56$~n\Eh\ deviation of $\epsilon^{(4)}$ reduces to $-10$~n\Eh\
for $\epsilon^{(6)}$.

A few comments regarding these observations are in order.
First of all, all deviations are smaller than the one-loop self-energy (and vacuum-polarization)
corrections known from nrQED \cite{Dr88,Pa06,PuKoCzPa16,PuKoPa17}, so both routes have the potential to provide a useful, quantitative description
of experimental observations. We have seen one indication that the differences 
appear to depend not only on the $Z$ nuclear charge number but also
on the electronic excitation in the system.
This connection is not so surprising, after all. From the no-pair variational aspect,
it is enough to remember that the generalization 
of Dirac's one-electron theory to poly-electron systems is challenging 
exactly because of the electron-electron interactions.

\clearpage

\begin{table}
  \caption{%
    Dirac--Coulomb (DC) and Dirac--Coulomb--Breit (DCB) electronic energy, in \Eh,
    obtained in the present no-pair variational framework.
    $\delta^{(4)}$ 
    and 
    $\delta^{(6)}$ 
    is the difference, in n\Eh, of the no-pair variational energy and 
    the leading and higher-order Foldy--Woythusen perturbation theory result (compiled 
    from the Refs.~\cite{PuKoPa17,PuKoCzPa16,Pa06,Dr88} or computed in this work), respectively.
    The nuclei are fixed near their equilibrium position at 1.4~bohr, 1.46~bohr, 1.65~bohr 
    in H$_2$, HeH$^+$, and in the equilateral triangular H$_3^+$, respectively.    
    \label{tab:all}
  }
  \begin{tabular}{@{}l lr lrr @{}}
    \hline\hline\\[-0.35cm]
    \multicolumn{1}{c}{} &	
    \multicolumn{1}{c}{$E_\DC$}  & $\delta^{(4)}_\DC$ &
    \multicolumn{1}{c}{$E_\DCB$} & $\delta^{(4)}_\DCB$ & $\delta^{(6)}_\DCB$ \\
    \cline{1-6}\\[-0.3cm]
     H$_2$   &  $-$1.174 489 754 & $\lbrace -21\rbrace$ &  $-$1.174 486 725(20) & $\lbrace -58\rbrace$ & $\lbrace-56\rbrace$ \\    
     He (1 $^1$S) &	$-$2.903 856 631   & $\lbrace -146\rbrace$ & $-$2.903 829 023(100) & $\lbrace -712\rbrace$ & $\lbrace -659 \rbrace$ \\
     He (2 $^1$S) & $-$2.146 084 791(3) & $\lbrace -22\rbrace$  & $-$2.146 082 424(3)  & $\lbrace -56 \rbrace$ & $\lbrace -10 \rbrace$ \\
     HeH$^+$ &	$-$2.978 834 635 & $\lbrace -142\rbrace$ &	$-$2.978 808 818(40) & $\lbrace -638 \rbrace$ & \\  
     H$_3^+$ &  $-$1.343 850 527(1) & $\lbrace -25\rbrace$ & $-$1.343 847 496(10) & $\lbrace -90\rbrace$ &\\              
    \cline{1-6}\\[-0.35cm]
    \cline{1-6}\\[-0.3cm]
  \end{tabular}
\end{table}

\clearpage
In summary, we have reported the development of a variational procedure for the no-pair Dirac--Coulomb--Breit Hamiltonian.
This procedure was used for atoms and molecules with clamped nuclei, currently with two, but straightforwardly extendable for more than two electrons, using explicitly correlated Gaussian functions and 
ultimately aiming at a parts-per-billion convergence of the energy. 
The procedure excellently reproduces literature data for two-electron atoms (ions) and the Dirac--Coulomb model.
Larger differences are observed with respect to Foldy--Woythusen perturbation theory
(FWPT, within the nrQED framework) already for atoms and molecules with low $Z$ values. 
Our variational DCB energies for the ground state of 
the H$_2$, H$_3^+$, and HeH$^+$ molecules and the He atom
are lower, by 
58-90~n\Eh
(for $Z=1$) and by 
638~n\Eh\ 
(for $Z=2$) than the FWPT energies.
These deviations are 1-2 orders of magnitude larger than the difference of the exact Dirac energy and the leading-order perturbation theory result for the one-electron hydrogen-like atoms 
(0.2 and 11~n\Eh\ for $Z=1$ and 2, respectively).
Higher-order ($m\alpha^6$) corrections to the FWPT (nrQED) energies, currently available for the H$_2$ molecule and the He atom, 
reduces the deviation a little bit, but do not change the order of magnitude of the difference.
The only exception in our test set is the excited 1s2s $^1$S state of the helium atom, for which the difference of the two approaches reduces to `only' 10~n\Eh\ when the higher-order corrections are also included in the FWPT energy (within the nrQED expansion).

At the same time, it is important to note that all listed deviations between the variational and FWPT energies are larger than the electron self energy predicted within the nrQED approach, 
hence our next priority is the calculation of this quantity for the present no-pair Dirac framework.
Furthermore, the effect of electron-positron pair creation (vacuum polarization) will be also accounted for. We are working
on the inclusion of the exact one-photon exchange to have
an improved description of the electron-electron interaction
beyond the zero-frequency approximation (Coulomb--Breit).

We think that the developed all-order, variational relativistic approach 
offers a broad perspective for further developments, 
we consider a possible inclusion of two- and multi-photon processes 
(including absorption and emission), 
either perturbatively or by an explicit account of the photon field in interaction with the fermionic degrees of freedom.

\vspace{0.5cm}
\begin{acknowledgments}
\noindent Financial support of the European Research Council through a Starting Grant (No.~851421) is gratefully acknowledged. 
\end{acknowledgments}


\end{document}